\documentclass[aps,prl,twocolumn,floats,showpacs,superscriptaddress,floatfix]{revtex4}
\usepackage{epsfig,natbib,float,amssymb}

\begin{document}
\title{Loss separation for dynamic hysteresis in magnetic thin films}
\author{Francesca Colaiori}
\affiliation{CNR-INFM, SMC, Dipartimento di Fisica, Universit\`a "La Sapienza", P.le A. Moro 2, 00185 Roma, Italy}
\author{Gianfranco Durin}
\affiliation{Istituto Nazionale di Ricerca Metrologica, 
strada delle Cacce 91, 10135 Torino, Italy}
\author{Stefano Zapperi}
\affiliation{CNR-INFM, SMC, Dipartimento di Fisica, Universit\`a "La Sapienza", P.le A. Moro 2, 00185 Roma, Italy}
\begin{abstract}
We develop a theory for dynamic hysteresis in ferromagnetic thin
films, on the basis of the phenomenological principle of loss
separation. We observe that, remarkably, the theory of loss
separation, originally derived for bulk metallic materials, is
applicable to disordered magnetic systems under fairly general
conditions regardless of the particular damping mechanism.  We confirm
our theory both by numerical simulations of a driven random--field
Ising model, and by re--examining several experimental data reported
in the literature on dynamic hysteresis in thin films. All the
experiments examined and the simulations find a natural interpretation
in terms of loss separation. The power losses dependence on the
driving field rate predicted by our theory fits satisfactorily all the
data in the entire frequency range, thus reconciling the apparent lack
of universality observed in different materials.
\end{abstract}
\maketitle 

Power losses in ferromagnetic materials generally depend on
the frequency of the applied field, a phenomenon referred to as
dynamic hysteresis \cite{CHA-99,MOO-04}.  The problem has important
implications for applications to high frequency devices, and, from a
purely theoretical point of view, for the understanding of the
dynamics of disordered magnetic systems, which represents a central
issue in non--equilibrium statistical mechanics.  While dynamic
hysteresis in metallic bulk three dimensional systems is well understood in
terms of eddy current dissipation, a satisfying theory for thin
films, where the effect of eddy current is expected to become 
negligible \cite{Bertotti}, is still lacking.  Hence, in recent years a great attention, both
experimental \cite{MOO-04,HE-93,LUS-94,JIA-95,RAQ-96,SUE-97,SUE-99,LEE-99,CHO-99,LEE-00,MOO-01,CHE-02,AST-02,SAN-03} and theoretical \cite{CHA-99,ZHO-95,ZHO-98,LYU-99}, has been devoted to
magnetic reversal in thin and ultrathin ferromagnetic films.

An accurate interpretation of the experimental data requires a
detailed understanding of the magnetization reversal properties on a
microscopic scale. Based on the analogy with Ising type models,
experimental data are often analyzed in terms of universal scaling
laws, such as the one relating the dynamical hysteresis loop area
($A$) to external parameters such temperature ($T$), amplitude
($H_0$), and frequency ($\omega$) of the applied magnetic field:
$A\propto H_0^{\alpha} \omega^{\beta}T^{-\gamma}$, where $\alpha$,
$\beta$, and $\gamma$ are scaling exponents \cite{CHA-99}.  The
experimental estimates of these exponents, often based on a very
limited scaling regime, display a huge variability
\cite{HE-93,LUS-94,JIA-95,RAQ-96,SUE-97,SUE-99,LEE-99,CHO-99,LEE-00},
so that the validity of a simple universal scaling law is still under
question.  Some authors also interpret the lack of good scaling of
$A(\omega)$ as a cross--over between two distinct dynamical regimes,
at low frequencies dominated by domain wall propagation, at high frequencies by the nucleation
of new domains \cite{MOO-04,LEE-99}. On the other hands, several authors recognized the present of a
static hysteretic term, yielding a non--vanishing loop area in the limit
$\omega\to 0$ \cite{ZHO-98,SAN-03,NIS-05}.  While this surely represents an important step forward for a better description of the data, we still need a more comprehensive theory to understand the magnetization reversal in thin films. 

For bulk ferromagnetic materials, the conventional theory of power losses 
takes as a starting point the phenomenological
principle of {\it loss separation} according to which the average
power loss can be decomposed into the sum of static (hysteretic),
classical and excess contributions: $P=P_{hyst}+P_{cl}+P_{ex}$
\cite{Bertotti}.  The hysteretic contribution $P_{hyst}$ is due to the
presence of quenched disorder, and gives rise to a non vanishing area
of the quasi--static hysteresis loops. The classical term $P_{cl}$ is
obtained computing the eddy currents present in an equivalent
homogeneously magnetized sample \cite{WIL-50}. Finally, excess losses
$P_{ex}$ take into account other contributions due to the dissipation
associated to the presence of multiple moving domain walls
\cite{BER-all}. The theory of loss separation
is usually considered inadequate for thin films, since eddy currents
dissipation disappears as the sample thickness goes to zero. While
this is certainly correct for what concerns the contribution from
classical losses, the hysteretic and excess loss contributions have been
 derived under
very general conditions which are in principle valid also for thin
films \cite{Bertotti}.

In this letter, we show that once the excess loss contribution is taken into
account, the theory of loss separation describes very accurately the
properties of dynamic hysteresis in thin films. 
Instead of directly considering power losses, we focus on the behavior
of the coercive field $H_c$ as a function of the driving field rate
$\dot{H}=dH/dt$, since these are the quantities used in published
experimental data. Analyzing the behavior of coercive
field is particularly convenient also because it is less sensitive
than the loop area to the particular driving condition employed in the
experiment (i.e. sinusoidal or triangular field, applied field or flux
rate, etc.). Moreover for fully saturated loops the two quantities are
proportional.

Analogously to the total losses, the coercive field can
be separated into three contributions
\begin{equation}
H_c=H_p+H_{cl}+H_{ex} ,
\label{eq:hc_loss}
\end{equation}
where $H_p$ is the static (hysteretic) contribution of the pinning field, $H_{cl}$
is the classical eddy current contribution, and $H_{ex}$ is the excess field. 

The classical term $H_{cl}=C_{cl} \dot{H}$ is linear in the driving
field rate $\dot{H}$ through a coefficient $C_{cl}\propto d^2\sigma
\mu$ \cite{Bertotti}, where $\sigma$ is the conductivity, $\mu$ is the
permeability, and $d$ is the sample thickness.  Clearly, this term goes rapidly
to zero with $d$, and thus it can be safely disregarded in thin films
and in the following we set $C_{cl}=0$.

The excess field has no explicit dependence on the thickness, thus in principle it should be taken into account also in thin films. Originally, it has been derived under two basic hypothesis: i) eddy currents provide the dominant dissipation mechanism of magnetization reversal, and ii) a number of correlated regions between \emph{magnetic objects} (domain walls, for instance) are progressively activated by the increase of the the magnetization rate $\dot{I}$. As a matter of fact, these hypothesis can be made more general. If we consider a single magnetic object which could reverse all the sample magnetization, the total excess field $H_{w}$ is taken proportional to the magnetization rate $\dot{I}$, so that $H_{w} = \Gamma \dot{I}$, where $\Gamma$ is an effective damping coefficient, This coefficient was originally calculated for the eddy current dissipation, but that can be generalized to \emph{any} general dissipation mechanism (spin relaxation, etc.) which produces an excess field $H_w$. Because in thin films the control parameter is $\dot{H}$, we also assume $\dot{I} = \mu \dot{H}$, where $\mu$ is the permeability. Given that, the factor $\Gamma \mu$ turns out to be an effective relaxation time relating the applied field rate to the excess field for a single magnetic object, so that  $H_{w} = \Gamma \mu \dot{H}$.

In the case of $n$ magnetic objects, the excess field $H_{ex}$ is proportionally reduced giving $H_{ex} = H_w / n$. At the same time, the number of active magnetic objects itself is a function of the excess field, and, as in the original model \cite{BER-all}, we assume a simple linear relation $n \simeq n_0+H_{exc}/V_0$, where $n_0$ is the number of active magnetic objects in the quasi--static limit, and $V_0$ is a characteristic field which controls the increase of the number of active magnetic objects due to the excess field. 

With these hypothesis, the excess field is calculated as 
\begin{equation}
H_{ex}= C_{ex}((1+r \dot{H})^{1/2}-1)
\label{eq:hc_ex}
\end{equation}

where $C_{ex}=n_0 V_0/2$, and $r = 4\Gamma \mu/(n_0^2 V_0)$. The excess field thus only depends on the typical relaxation time $\Gamma \mu$, the number of magnetic objects in the quasi--static limit $n_0$, and the characteristic field $V_0$. Remarkably, as reported in \cite{Bertotti}, when $r \dot{H} \ll 1$, we get $H_{ex} = \Gamma \mu \dot{H} /n_0$, that is, for a large number of magnetic objects ($n \sim n_0 \gg 1$) the excess field is a linear function of the applied field rate. On the other hand, for $r \dot{H} \gg 1$, $H_{ex} = \sqrt{\Gamma \mu V_0 \dot{H}}$, i.e. the excess field has a square-root dependence on the rate when the number of magnetic objects is pretty small ($n_0 \sim 1$).

\begin{figure}
\centerline{\psfig{file=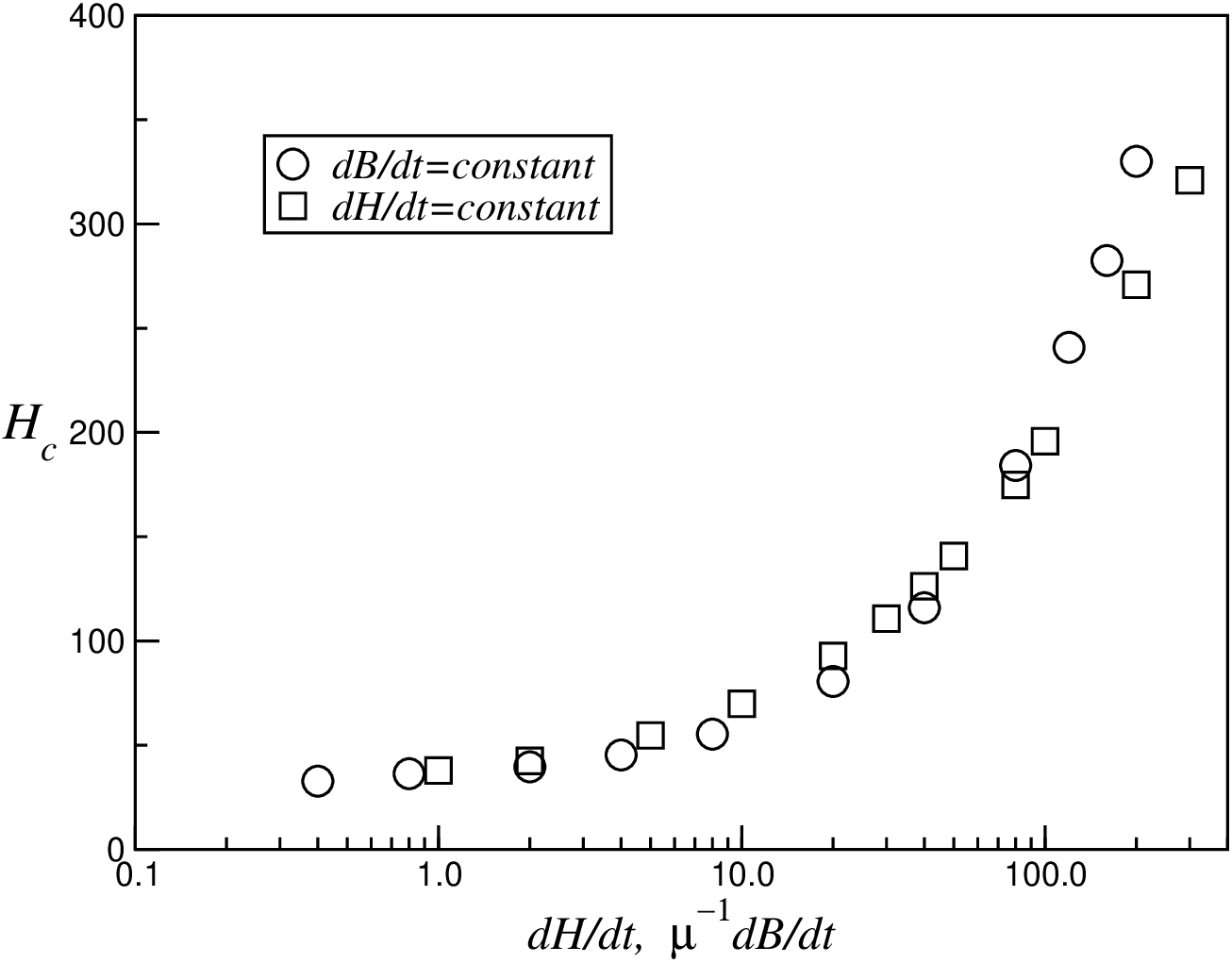,width=8cm}}
\caption{Variation of dynamic coercive field $H_c$ with the applied field rate $dH/dt$, for a Fe-based amorphous ribbon with thickness 20 $\mu m$. The coercive field measured under constant field rate $\dot{H}$ (squares) and constant magnetization rate $\dot{I}$ (circles) collapses using $\dot{H} = \dot{I}/\mu$, where  $\mu= 0.08$ is the linear permeability around the coercive field.}
\label{fig:hc}
\end{figure}

At this point, one can argue about the equivalence of performing experiments by controlling the driving field $\dot{H}$ instead of the magnetization rate $\dot{I}$ in dynamic experiments.  To verify this assumption, we have performed dynamic hysteresis experiments in an $Fe_{64}Co_{21}B_{15}$ amorphous ribbon having a thickness of 20 $\mu m$, a typical sample where the loss separation is known to apply. We have measured the coercive field both as a function of $\dot{H}$ and of $\dot{I}$.  As shown in Fig.~\ref{fig:hc}, the data from the two experiments collapse reasonably well using the linear permeability $\mu \simeq 0.08$ measured around the coercive field.

\begin{figure}
\centerline{\psfig{file=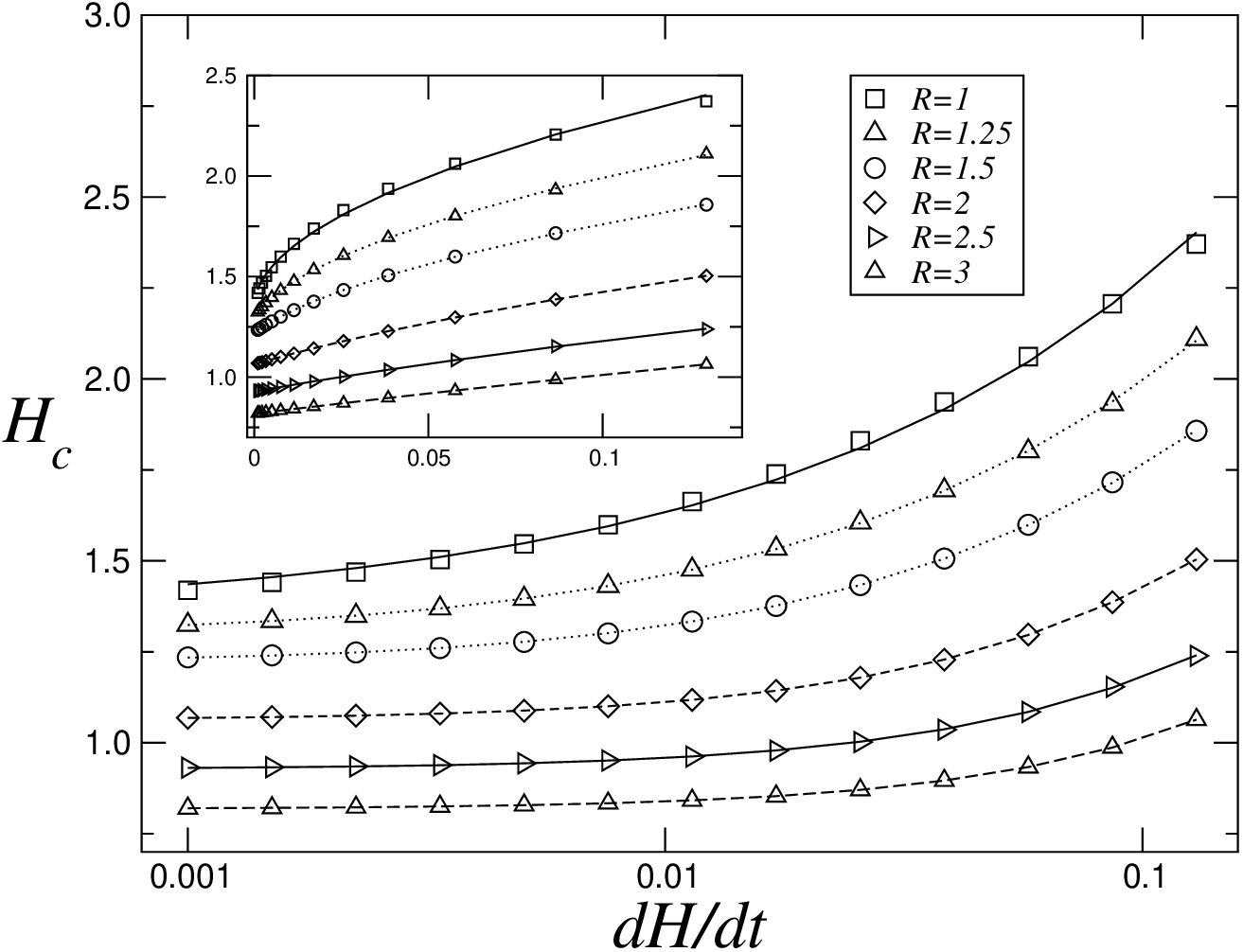,width=8cm}}
\caption{The dynamic coercive field as a function of the applied
field rate from simulations of the driven RFIM with different values
of the variance $R$ of the disorder distribution.  The
curves are fitted according to Eq.~\protect\ref{eq:hc_ex}, with
$C_{cl}=0$.  In the inset: data in lin-lin scale. Note the linear behavior at large $R$, and 
the square-root dependence at low $R$.}
\label{fig:rfim1}
\end{figure}

In order to test the generality of the loss separation mechanisms for
hysteresis we consider the driven random-field Ising model (RFIM),
which has been proposed in the past as a paradigmatic model to
understand the effect of disorder in ferromagnetic hysteresis
\cite{SET-93,DAH-96,PER-99}. In the RFIM, a spin $s_i = \pm 1$ is
assigned to each site of two dimensional square lattice.  The spins
are coupled to their nearest--neighbors spins by a ferromagnetic
interaction of strength $J$ and to the external field $H$. In
addition, to each site is associated a random field $h_i$ taken from a
Gaussian probability density $\rho(h)=\exp(-h^2/2R^2)/\sqrt{2\pi}R$,
with width $R$, which measures the strength of the disorder. The
Hamiltonian thus reads
\begin{equation} 
{\cal H} = -\sum_{\langle i,j \rangle}Js_i s_j -\sum_i(H+h_i)s_i \,,
\label{eq:rfim}
\end{equation}
where $\sum_{\langle i,j \rangle}$ is restricted to nearest-neighbors
pairs.  We consider a simple relaxation dynamics
obtained in the limit $T \rightarrow 0$ of the Glauber dynamics
\cite{SET-93,DAH-96,PER-99}: at each time step the spins align with
the local effective field $s_i = \mbox{sign}\left(J\sum_j s_j  + h_i +H\right)$.

This model was originally studied in the quasistatic limit, however,
dynamic effects can easily be incorporated assigning a timescale
$\tau$ to spin relaxation and increasing the magnetic field in steps
$\Delta$. This is equivalent to an effective field rate $\dot{H}\equiv\Delta/\tau$.
We numerically compute hysteresis loops for different values of the
disorder strength $R$ and obtain the coercive field averaging over
different realizations of the disorder. In Fig.~\ref{fig:rfim1}, we
show that the data are described well by Eqs.\ref{eq:hc_loss}--\ref{eq:hc_ex}. The
resulting fit indicates that $C_{ex}$ grows with the disorder, while
$r$ decreases. This fact can be understood by noting that the growth of
disorder naturally leads to an increase in the number of the domains.  In fact, when $R \sim 1$, $n_0$ is pretty small so that $r \dot{H} \gg 1$ and the excess field has a square-root dependence on the rate $\dot{H}$. At $R \sim 3$, $n_0$ is large ($r \dot{H} \lesssim 1$) and the dependence is linear (see the inset of
Fig.~\ref{fig:rfim1}).

\begin{figure}[ht]\vspace{0.5cm}
 \includegraphics[width=8cm]{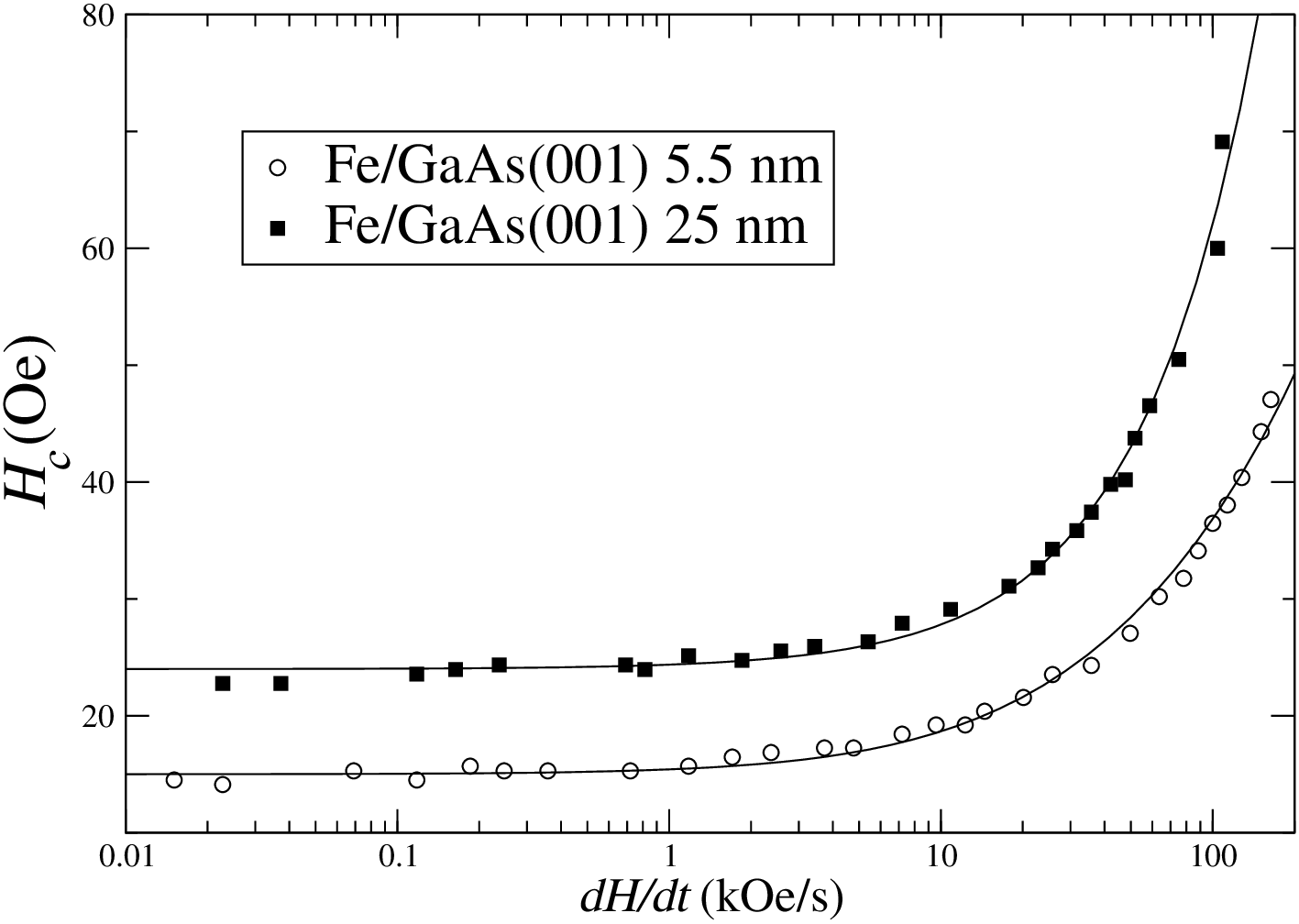}
\caption{Analysis of experimental data using 
Eqs.~\protect\ref{eq:hc_loss}--\ref{eq:hc_ex} on epitaxial Fe/GaAs(001) thin films, as reported
in Ref.~\protect\cite{LEE-99}. Data for the 25 nm film ($\blacksquare$) are approximately described by a linear function}
\label{fig:FeGaAs}
\end{figure}

\begin{figure}[ht]\vspace{0.5cm}
 \includegraphics[width=8cm]{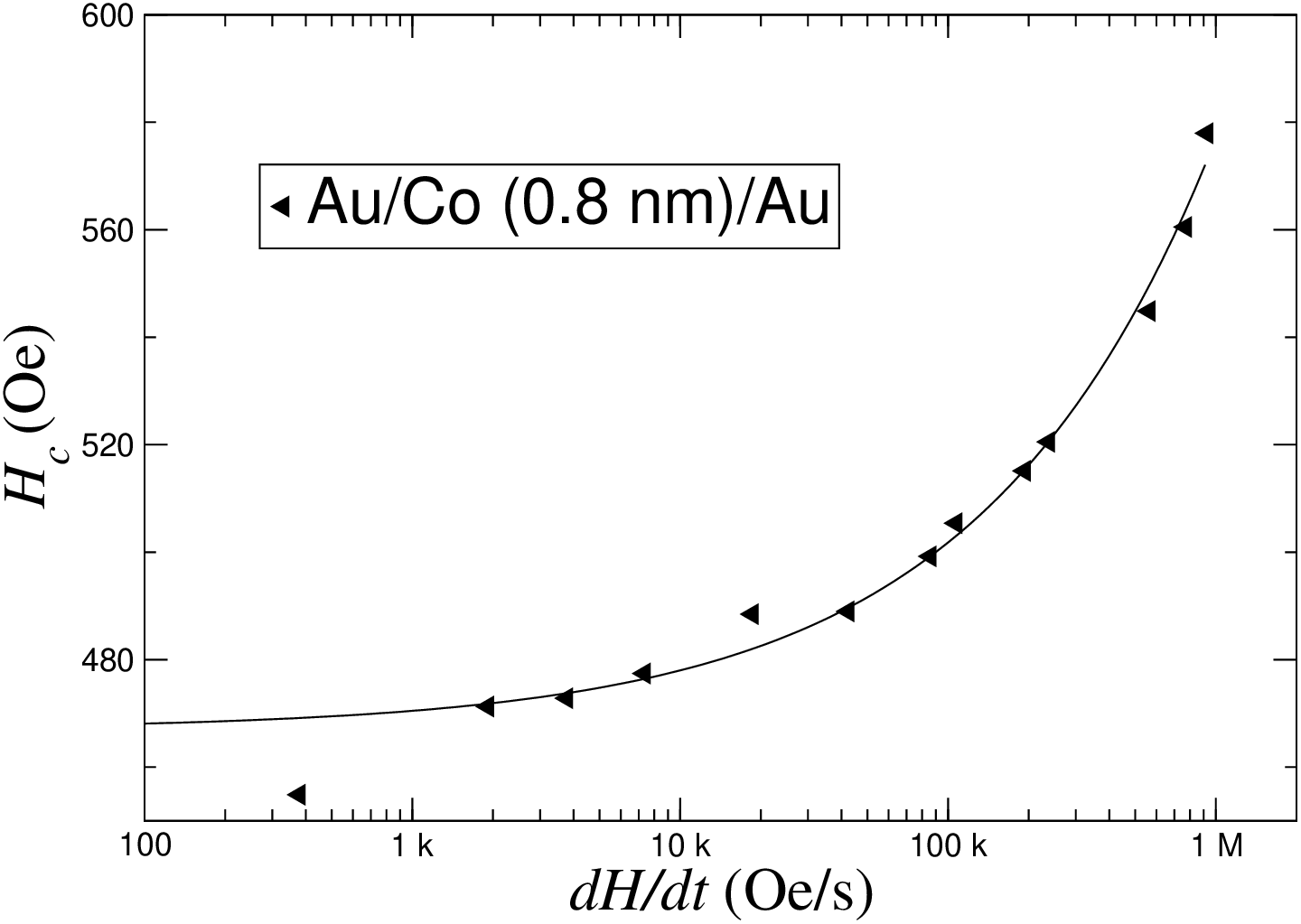}
\caption{Analysis of experimental data using 
Eqs.~\protect\ref{eq:hc_loss}--\ref{eq:hc_ex} on a MoS$_2$/Au/Co/Au sandwich taken at 300 K, as reported in Ref.~\protect\cite{RAQ-96}. The excess field has approximately a square-root dependence on the applied field rate.}
\label{fig:AuCoAu}
\end{figure}

It is interesting to remark that in
Ref.~\cite{ZHE-02c} similar results for the loop area in the RFIM have
been fitted by a scaling law $A=A_0+A_1 \omega^\beta$, obtaining a
disorder dependent $\beta$.  While it is difficult to discriminate
between different fitting functions, we notice that an exponent
depending on microscopic parameters is difficult to justify
theoretically.  In this respect, Eq.~\ref{eq:hc_loss} interpolates
between $\beta=1$ for $ r \dot{H} \ll 1$ and $\beta=1/2$ for $r \dot{H}
\gg 1$, as said.  We also note that $\beta=1/2$ was obtained from the solution of a model for
the dynamics of a single domain wall \cite{SAN-03,RUI-02}, fully compatible with the present results.

\begin{figure}[h]\vspace{0.5cm}
 \includegraphics[width=8cm]{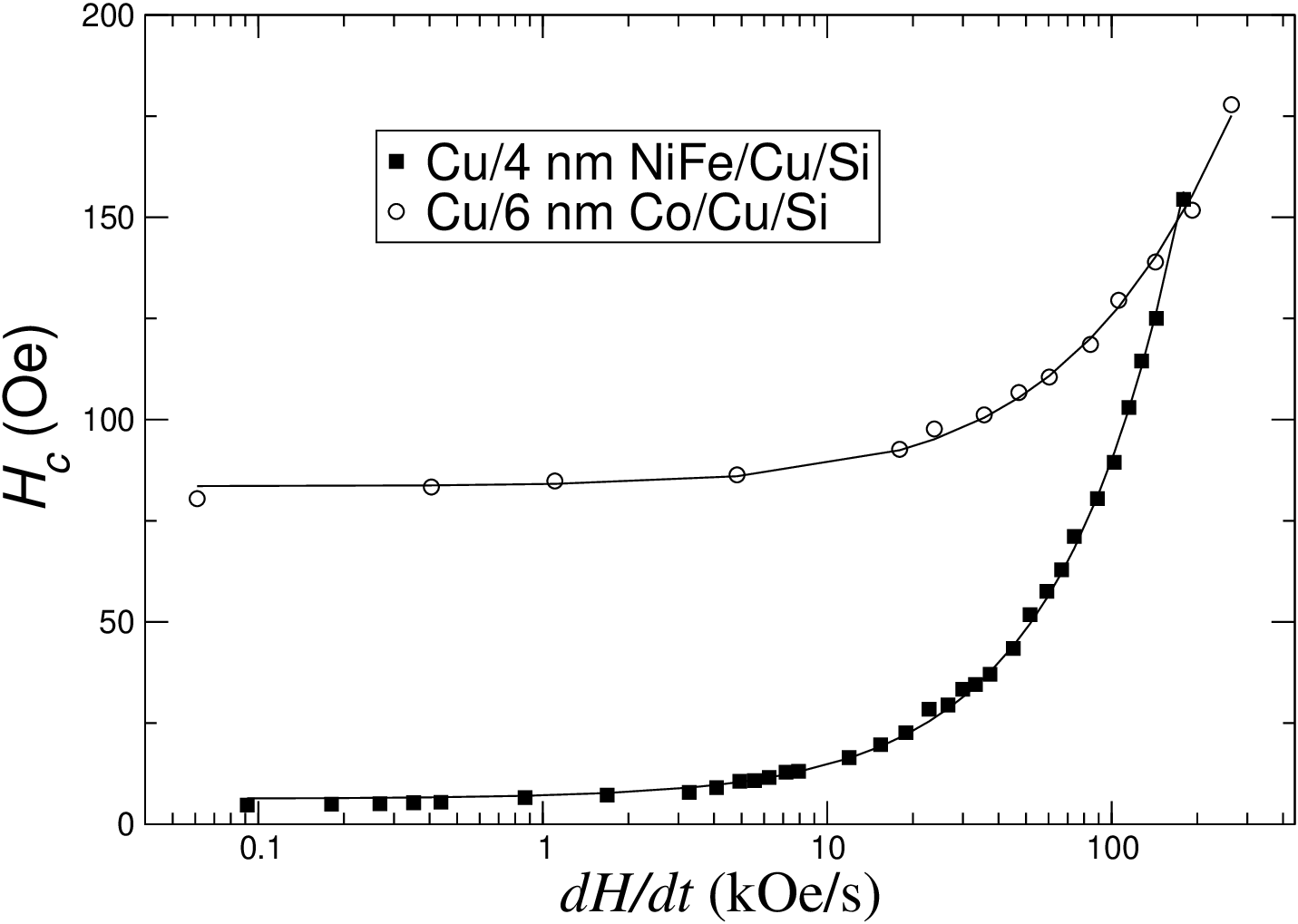}
\caption{Analysis of experimental data using 
Eqs.~\protect\ref{eq:hc_loss}--\ref{eq:hc_ex} on a NiFe and Co single magnetic layer films, as reported
in Ref.~\protect\cite{LEE-00}. Data for the NiFe ($\blacksquare$) are approximately described by a linear function.}
\label{fig:CuFMCuSi}
\end{figure}

\begin{figure}[ht]\vspace{0.5cm}
 \includegraphics[width=8cm]{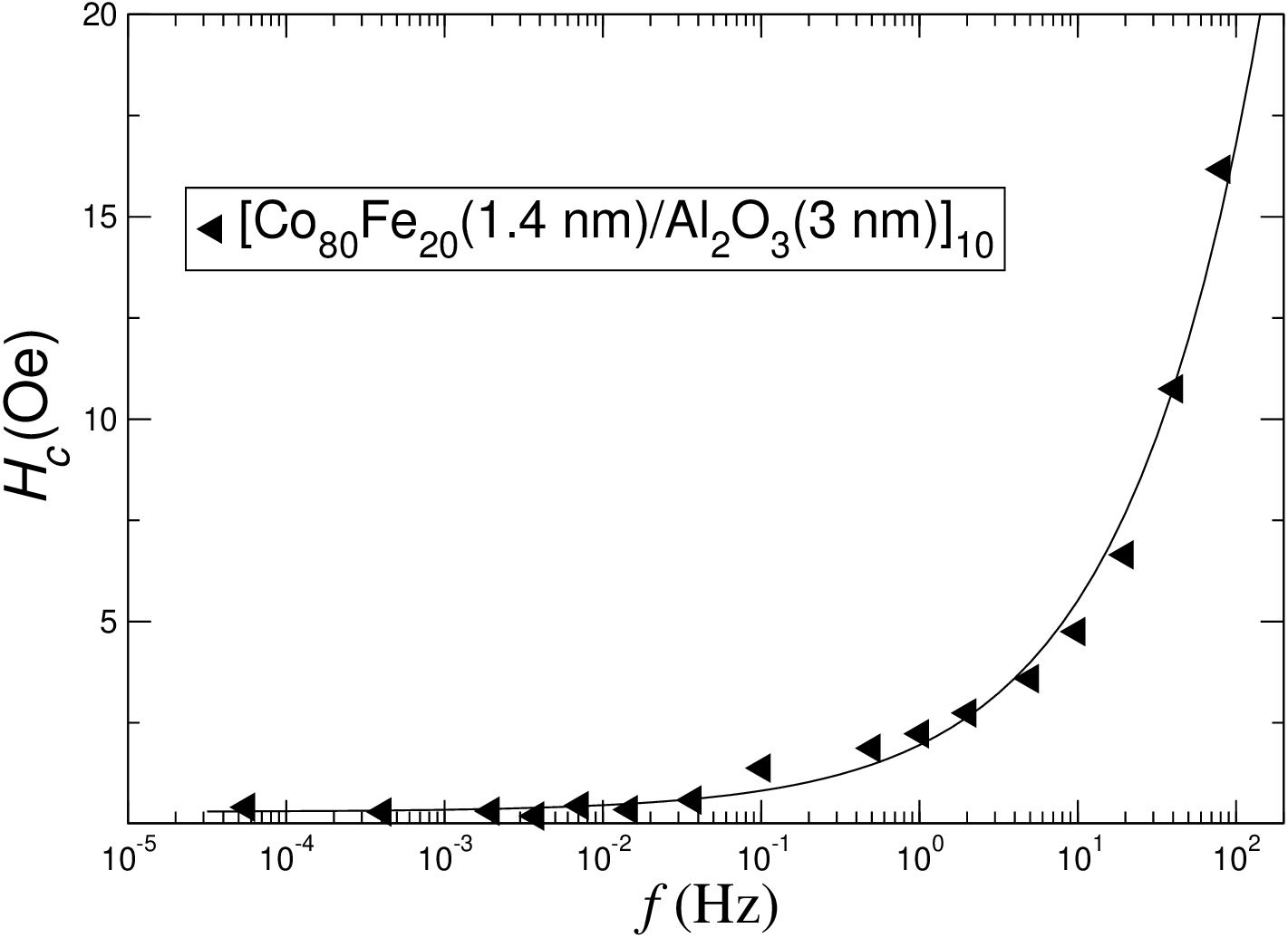}
\caption{Analysis of experimental data using Eqs.~\protect\ref{eq:hc_loss}--\ref{eq:hc_ex} on a superparamagnetic multilayer [Co$_{80}$Fe$_{20}$(1.4 nm)/Al$_{2}$O$_{3}$ (3 nm)]$_{10}$, as reported in Ref.~\protect\cite{CHE-02}. The excess field has approximately a square-root dependence on the applied field rate.}
\label{fig:CoFe}
\end{figure}

As Eqs.~\ref{eq:hc_loss}--\ref{eq:hc_ex} provide a good description of dynamic
hysteresis under fairly general conditions, we can try to compare its
predictions with some of the experimental measurements reported in the
literature. In particular, we re--examine the measurements of dynamic
coercivity in magnetic thin films and multilayers of various thicknesses, as reported in Figs.~\ref{fig:FeGaAs}--\ref{fig:CoFe} (from Refs.~\cite{LEE-99,RAQ-96,LEE-00,CHE-02}, respectively).
Significantly, all the data are well fitted by Eqs.~\ref{eq:hc_loss}--\ref{eq:hc_ex} over the entire range of the applied field rate. Both limiting cases are present ($r \dot{H} \ll 1$, and $r \dot{H} \gg 1$), with an approximately linear (black squares) and a square-root (black triangles) dependence on $\dot{H}$. Despite the large variability of compositions, structures, and static coercive fields (which span two orders of magnitude), the fitting curves are compatible with a limited range of the model parameters: $n_0 = 1-100$, $V_0 = 0.1-10$ Oe, and $\Gamma \mu = 10^{-2}-10^{-4} s$. In particular, the time costant lays within the expected range for the dynamics of domain walls in thin films \cite{MOO-04}. This observation strongly supports the validity and the large applicability of this model. This also seriously puts in doubt the existence of a dynamic transition between a regime dominated by propagation and one dominated by nucleation of new domains. As a matter of fact, the model does not exclude the nucleation of domains, as an increasing value of $n$ can imply both the activation of existing magnetic objects and the creations of new ones. But we also must note that in the case of a square-root dependence on the rate, the number of magnetic objects cannot change significantly, and that also in the opposite case of a linear dependence, when the number of static magnetic objects is large the effect of nucleation of new active domains is practically irrelevant.

In summary we show that the properties of dynamic hysteresis in
ferromagnetic thin films can be explained in terms of the theory 
of loss separation, originally derived for bulk metallic materials.
This conclusion is supported by numerical simulations and by the analysis
of experimental data reported in the literature.

\end{document}